\documentclass{ws-p8-50x6-00}
\usepackage{graphicx}
\usepackage{amsmath}
\usepackage{amsfonts}
\usepackage{amssymb}

\begin{document}

\title{Tutorial on quasi-sparse eigenvector diagonalization}
\author{Dean Lee}
\address{University of Massachusetts, Amherst, MA 01003  USA
\\ dlee@physics.umass.edu}
\maketitle
\abstracts{
We review several topics related to the diagonalization of quantum
field Hamiltonians using the quasi-sparse eigenvector (QSE) method.}

\section{Introduction}

Quasi-sparse eigenvector (QSE) diagonalization is a new computational method
which finds the low-lying eigenvalues and eigenvectors for a general quantum
field Hamiltonian \cite{qse}. \ It is able to handle the exponential increase
in the size of Fock space for large systems by exploiting the sparsity of the
Hamiltonian. \ QSE diagonalization can even be applied
directly to infinite-dimensional systems. \ The method is most effective when
the splitting between low-lying eigenvalues is not too small compared to the
size of the off-diagonal Hamiltonian matrix entries. \ In such cases the low-lying
eigenvectors are quasi-sparse, meaning that the vector is dominated by a small
fraction of its largest components. \ The QSE algorithm can then be applied to 
the Hamiltonian $H$ using the following steps:

\begin{enumerate}
\item  Select a subset of orthonormal basis vectors $\left\{  e_{i_{1}}%
,\cdots,e_{i_{n}}\right\}  $ and call the corresponding subspace $S$.

\item  Diagonalize $H$ restricted to $S$ and find one eigenvector $v.$

\item  Sort the basis components $\langle e_{i_{j}}|v\rangle$ according to
their magnitude and remove the least important basis vectors.

\item  Replace the discarded basis vectors by new basis vectors. \ These are
selected at random from a pool of candidate basis vectors which are connected
to the old basis vectors through non-vanishing matrix elements of $H$.

\item  Redefine $S$ as the subspace spanned by the updated set of basis
vectors and repeat steps 2 through 5.
\end{enumerate}

\noindent If the subset of basis vectors is sufficiently large, the exact
eigenvectors will be stable fixed points of the update process.

The purpose of this short tutorial is to provide additional information for
those interested in writing or using codes that implement QSE methods. \ We
discuss the calculation of Hamiltonian matrix elements in the momentum Fock
space representation and the generation and selection of new basis states.
\ We also include a simple program which applies the QSE method to find the
ground state of an input matrix. \ Readers interested in an introductory
overview of QSE diagonalization are encouraged to consult the references in
\cite{qse}.

\section{Fock states and the Hamiltonian matrix}

It is most effective to describe the details of a QSE calculation in the
context of a specific example. \ In this discussion we consider $\phi^{4}$
theory in $1+1$ dimensions, the same example as in \cite{qse}. \ Some of the
techniques described here have been specifically optimized for scalar field
theories. \ Other systems require a somewhat different set of tools and will
be discussed in the future.

The Hamiltonian density for $\phi^{4}$ theory in $1+1$ dimensions has the form
\[
\mathcal{H}=\tfrac{1}{2}\left(  \tfrac{\partial\phi}{\partial t}\right)
^{2}+\tfrac{1}{2}\left(  \tfrac{\partial\phi}{\partial x}\right)  ^{2}%
+\tfrac{\mu^{2}}{2}\phi^{2}+\tfrac{\lambda}{4!}\text{:}\phi^{4}\text{:},
\]
where the normal ordering is with respect to the mass $\mu$. \ We consider the
system in a periodic box of length $2L$. \ We then expand in momentum modes
and reinterpret\ the problem as an equivalent Schr\"{o}dinger equation
\cite{periodic}. We will use a momentum cutoff scheme where the modes $q_{n}$ 
are restricted in momentum so that $\left| n \right| \leq N_{max}$.
\ The resulting Hamiltonian is
\begin{align}
H  &  =-\tfrac{1}{2}%
{\displaystyle\sum_{n}}
\tfrac{\partial}{\partial q_{-n}}\tfrac{\partial}{\partial q_{n}}+\tfrac{1}{2}%
{\displaystyle\sum_{n}}
\left(  \omega_{n}^{2}(\mu)-\tfrac{\lambda b(\mu)}{4L}\right)  \,q_{-n}%
q_{n}\label{h}\\
&  +\tfrac{\lambda}{4!2L}%
{\displaystyle\sum_{n_{1}+n_{2}+n_{3}+n_{4}=0}}
q_{n_{1}}q_{n_{2}}q_{n_{3}}q_{n_{4}},\nonumber
\end{align}
where
\begin{equation}
\omega_{n}(\mu)=\sqrt{\tfrac{n^{2}\pi^{2}}{L^{2}}+\mu^{2}}
\end{equation}
and $b(\mu)$ is the coefficient for the mass counterterm
\begin{equation}
b(\mu)=%
{\displaystyle\sum_{n}}
\tfrac{1}{2\omega_{n}(\mu)}. \label{c}%
\end{equation}

It is convenient to split the Hamiltonian into free and interacting parts with
respect to an arbitrary mass $\mu^{\prime}$:%

\begin{equation}
H_{free}=-\tfrac{1}{2}%
{\displaystyle\sum_{n}}
\tfrac{\partial}{\partial q_{-n}}\tfrac{\partial}{\partial q_{n}}+\tfrac{1}{2}%
{\displaystyle\sum_{n}}
\omega_{n}^{2}(\mu^{\prime})\,q_{-n}q_{n},
\end{equation}%
\begin{align}
H  &  =H_{free}+\tfrac{1}{2}%
{\displaystyle\sum_{n}}
\left(  \mu^{2}-\mu^{\prime2}-\tfrac{\lambda b(\mu)}{4L}\right)  q_{-n}q_{n}\\
&  +\tfrac{\lambda}{4!2L}%
{\displaystyle\sum_{n_{1}+n_{2}+n_{3}+n_{4}=0}}
q_{n_{1}}q_{n_{2}}q_{n_{3}}q_{n_{4}}.\nonumber
\end{align}
$\mu^{\prime}$ will be used to define the basis states of our Fock space.
\ Since $H$ is independent $\mu^{\prime}$, it is useful to perform
calculations for different $\mu^{\prime}$ to obtain an estimate of the
error. \ It is also helpful to find the range of values for $\mu^{\prime}$
which maximizes the quasi-sparsity of the eigenvectors. \ This can be
used to improve the accuracy of the QSE calculation.

Let us define ladder operators with respect to $\mu^{\prime}$,
\begin{align}
a_{n}(\mu^{\prime})  &  =\tfrac{1}{\sqrt{2\omega_{n}(\mu^{\prime})}%
}\left[  q_{n}\omega_{n}(\mu^{\prime})+\tfrac{\partial}{\partial q_{-n}%
}\right] \\
a_{n}^{\dagger}(\mu^{\prime})  &  =\tfrac{1}{\sqrt{2\omega_{n}%
(\mu^{\prime})}}\left[  q_{-n}\omega_{n}(\mu^{\prime})-\tfrac{\partial
}{\partial q_{n}}\right]  .
\end{align}
These satisfy the usual commutation relations,%

\begin{align}
\left[  a_{n},a_{m}^{\dagger}\right]   &  =\delta_{nm}\\
\left[  a_{n},a_{m}\right]   &  =\left[  a_{n}^{\dagger},a_{m}^{\dagger
}\right]  =0.
\end{align}
The Hamiltonian can now be rewritten as%

\begin{align}
H  &  =%
{\displaystyle\sum_{n}}
\omega_{n}(\mu^{\prime})a_{n}^{\dagger}a_{n}+\tfrac{1}{4}(\mu^{2}-\mu
^{\prime2}-\tfrac{\lambda b(\mu)}{4L})%
{\displaystyle\sum_{n}}
\tfrac{\left(  a_{-n}+a_{n}^{\dagger}\right)  \left(  a_{n}+a_{-n}^{\dagger
}\right)  }{\omega_{n}(\mu^{\prime})}\label{ha}\\
&  +\tfrac{\lambda}{192L}%
{\displaystyle\sum_{n_{1}+n_{2}+n_{3}+n_{4}=0}}
\left[  \tfrac{\left(  a_{n_{1}}+a_{-n_{1}}^{\dagger}\right)  }{\sqrt
{\omega_{n_{1}}(\mu^{\prime})}}\tfrac{\left(  a_{n_{2}}+a_{-n_{2}}^{\dagger
}\right)  }{\sqrt{\omega_{n_{2}}(\mu^{\prime})}}\tfrac{\left(  a_{n_{3}%
}+a_{-n_{3}}^{\dagger}\right)  }{\sqrt{\omega_{n_{3}}(\mu^{\prime})}}%
\tfrac{\left(  a_{n_{4}}+a_{-n_{4}}^{\dagger}\right)  }{\sqrt{\omega_{n_{4}%
}(\mu^{\prime})}}\right]  .\nonumber
\end{align}
In (\ref{ha}) we have omitted constants contributing only to the vacuum energy.

We can represent any momentum-space Fock state as a string of occupation
numbers,
\begin{equation}
\left|  o_{-N_{\max}},\cdots,o_{N_{\max}}\right\rangle ,
\end{equation}
where \
\begin{equation}
a_{n}^{\dagger}a_{n}\left|  o_{-N_{\max}},\cdots,o_{N_{\max}}\right\rangle
=o_{n}\left|  o_{-N_{\max}},\cdots,o_{N_{\max}}\right\rangle .
\end{equation}
From the usual ladder operator relations, it is straightforward to calculate
the matrix element of $H$ between two arbitrary Fock states. \ In \cite{qse}
auxiliary cutoffs were used to render the dimension of Fock space finite.
\ This is in fact an unnecessary restriction and QSE diagonalization
encounters no difficulties dealing with the full infinite-dimensional space.

\section{New basis vectors}

Aside from calculating matrix elements, the only other fundamental operation
involving basis vectors is the generation of new basis vectors. \ As mentioned
before, the new states should be connected to an old basis vector through
non-vanishing matrix elements of $H$. \ Let us refer to the old basis vector
as $\left|  e\right\rangle $. \ For this example there are two types of terms
in our interaction Hamiltonian, a quartic interaction
\begin{equation}%
{\displaystyle\sum_{n_{1},n_{2},n_{3}}}
\left(  a_{n_{1}}+a_{-n_{1}}^{\dagger}\right)  \left(  a_{n_{2}}+a_{-n_{2}%
}^{\dagger}\right)  \left(  a_{n_{3}}+a_{-n_{3}}^{\dagger}\right)  \left(
a_{-n_{1}-n_{2}-n_{3}}+a_{n_{1}+n_{2}+n_{3}}^{\dagger}\right)  ,
\end{equation}
and a quadratic interaction
\begin{equation}%
{\displaystyle\sum_{n}}
\left(  a_{-n}+a_{n}^{\dagger}\right)  \left(  a_{n}+a_{-n}^{\dagger}\right)
.
\end{equation}
To produce a new vector from $\left|  e\right\rangle $ we simply choose one of
the possible operator monomials%
\begin{align}
&  a_{n_{1}}a_{n_{2}}a_{n_{3}}a_{-n_{1}-n_{2}-n_{3}},\,a_{-n_{1}}^{\dagger
}a_{n_{2}}a_{n_{3}}a_{-n_{1}-n_{2}-n_{3}},\cdots,\\
&  a_{-n}a_{n},\,a_{n}^{\dagger}a_{-n}^{\dagger},\cdots\nonumber
\end{align}
and let it act upon $\left|  e\right\rangle $. \ Our experience is that the
interactions involving the small momentum modes are generally more important
than those for the large momentum modes, a signal that the ultraviolet
divergences have been properly renormalized. \ For this reason it is best to
arrange the selection probabilities such that the smaller values of $\left|
n_{1}\right|  $, $\left|  n_{2}\right|  $, $\left|  n_{3}\right|  $ and
$\left|  n\right|  $ are chosen more often.

We note that the new basis vector selection will occasionally fail. \ Either the vector is
zero due to an annihilation operator acting on an unoccupied state or the
vector is already in our subset of basis vectors and therefore not new. \ In
either case we simply select a new basis vector again. \ If the
selection process fails repeatedly then a different old basis vector $\left|
e^{\prime}\right\rangle $ is used.

\section{Sample code}

We have considered two basic operations, the calculation of matrix elements
and the generation of new basis states. \ The subroutines which perform these
tasks will depend on the form of the quantum field Hamiltonian. \ The main
structure of the QSE algorithm, however, is independent of the details of the
Hamiltonian. \ We demonstrate the essential features of the algorithm with the
following short MATLAB program. \ In this example we find the ground state of
a finite matrix. To keep the example as simple as possible, we avoid
calculating matrix elements and generating new basis states by assuming
that the entire Hamiltonian matrix can be stored in memory.\medskip

\qquad\texttt{niter = 30; }

\qquad\texttt{nactive = 100; }

\qquad\texttt{nretain = 80; }

\qquad\texttt{H = loadsparse('samplematrix'); }

\qquad\texttt{vecs = [1:nactive]; }

\qquad\texttt{for iter = 1:niter}

\qquad\qquad\texttt{\lbrack v,d] = eigs(H(vecs,vecs),'SR',1);}

\qquad\qquad\texttt{weights1 = v.\symbol{94}2; }

\qquad\qquad\texttt{\lbrack sortelements, sortorder] = sort(-weights1); }

\qquad\qquad\texttt{vecs = vecs(sortorder(1:nretain)); }

\qquad\qquad\texttt{wtsum1 = cumsum(weights1)/sum(weights1); }

\qquad\qquad\texttt{while (length(vecs)
$<$%
nactive) }

\qquad\qquad\qquad\texttt{over1 = find(rand*wtsum1(nretain)
$<$%
wtsum1); }

\qquad\qquad\qquad\texttt{seedvec = vecs(over1(1)); }

\qquad\qquad\qquad\texttt{candidates = find(H(:,seedvec)); }

\qquad\qquad\qquad\texttt{candidates = candidates(candidates \symbol{126}=
seedvec); }

\qquad\qquad\qquad\texttt{weights2 = abs(H(candidates,seedvec)); }

\qquad\qquad\qquad\texttt{wtsum2 = cumsum(weights2)/sum(weights2); }

\qquad\qquad\qquad\texttt{over2 = find(rand
$<$%
wtsum2); }

\qquad\qquad\qquad\texttt{newvec = candidates(over2(1)); }

\qquad\qquad\qquad\texttt{vecs = unique([vecs newvec]); }

\qquad\qquad\texttt{end }

\qquad\texttt{end}

\qquad\texttt{save results.mat vecs v d}\medskip

\texttt{niter} is the number of iterations of the algorithm.
\ \texttt{nactive} is the number of basis states used to form the subspace
$S$, and \texttt{nretain} is the number of basis states kept after removing
the tail of the eigenvector distribution. \ \texttt{loadsparse} is a program
which reads in a datafile called \texttt{samplematrix}. \ This data is stored
in the matrix variable \texttt{H}. \ The remaining code can be broken down in
terms of the five basic steps of the QSE algorithm.\medskip

\noindent\textit{Select a subset of orthonormal basis vectors }$\left\{
e_{i_{1}},\cdots,e_{i_{n}}\right\}  $\textit{ and call the corresponding
subspace }$S$\textit{.}

\noindent\qquad\texttt{vecs = [1:nactive];}

\noindent The variable \texttt{vecs} is a row vector which labels the basis
vectors in the subspace $S$. \ In our example \texttt{vecs} is initialized to
be the first \texttt{nactive} basis vectors. \ We are assuming that the ground
state of the system has a non-zero overlap with at least one of the first
\texttt{nactive} basis vectors. \ A more general initialization would be to choose a
random subset of basis vectors.\medskip

\noindent\textit{Diagonalize }$H$\textit{ restricted to }$S$\textit{ and find
one eigenvector }$v.$

\noindent\qquad\texttt{\lbrack v,d] = eigs(H(vecs,vecs),'SR',1);}

\noindent The function \texttt{eigs} is a sparse matrix Arnoldi
diagonalization routine. \ It is being called with the parameters
\texttt{'SR',1} which tells \texttt{eigs} to find the eigenvalue and
eigenvector with the smallest real part.  The Hamiltonian submatrix corresponding with basis vectors in \texttt{vecs} is diagonalized.  \ The eigenvector is stored in the
column vector \texttt{v} and the eigenvalue is stored as the variable
\texttt{d}.\medskip

\noindent\textit{Sort the basis components }$\langle e_{i_{j}}|v\rangle
$\textit{ according to their magnitude and remove the least important basis vectors.}

\noindent\qquad\texttt{weights1 = v.\symbol{94}2;}

\noindent\qquad\texttt{\lbrack sortelements, sortorder] = sort(-weights1); }

\noindent\qquad\texttt{vecs = vecs(sortorder(1:nretain));}

\noindent \texttt{weights1} is a column vector which contains the squares
of the components of the eigenvector \texttt{v}$.$ \ \texttt{sortelements} is
a list of the elements of \texttt{-weights1} from smallest to greatest.
\ \texttt{sortorder} is the corresponding list of locations for these
elements. \ If for example \texttt{weights1 = [.2 .5 .3]}, then
\texttt{sortorder = [2 3 1]}. \ The list \texttt{sortelements} is not used.
\ Only the \texttt{nretain} most important basis vectors are kept in
\texttt{vecs}.\texttt{\medskip}

\noindent\textit{Replace the discarded basis vectors by new basis vectors.
\ These are selected at random from a pool of candidate basis vectors which
are connected to the old basis vectors through non-vanishing matrix elements
of }$H$\textit{.}

\noindent\qquad\texttt{wtsum1 = cumsum(weights1)/sum(weights1); }

\noindent\qquad\texttt{while (length(vecs)
$<$%
nactive) }

\noindent\qquad\qquad\texttt{over1 = find(rand*wtsum1(nretain)
$<$%
wtsum1); }

\noindent\qquad\qquad\texttt{seedvec = vecs(over1(1)); }

\noindent\qquad\qquad\texttt{candidates = find(H(:,seedvec)); }

\noindent\qquad\qquad\texttt{candidates = candidates(candidates \symbol{126}=
seedvec); }

\noindent\qquad\qquad\texttt{weights2 = abs(H(candidates,seedvec)); }

\noindent\qquad\qquad\texttt{wtsum2 = cumsum(weights2)/sum(weights2); }

\noindent\qquad\qquad\texttt{over2 = find(rand
$<$%
wtsum2); }

\noindent\qquad\qquad\texttt{newvec = candidates(over2(1)); }

\noindent\qquad\qquad\texttt{vecs = unique([vecs newvec]); }

\noindent\qquad\texttt{end}

\noindent\texttt{cumsum} is a function which takes the input list and creates
another list consisting of partial sums. \ \texttt{wtsum1} is therefore an
ascending string of numbers between $0$ and 1 such that the difference between
an entry and its preceding entry is proportional to the corresponding value of
\texttt{weights1}. \ \texttt{over1} is a location list of elements of
\texttt{wtsum1} which are greater than the product of a generated random
number between 0 and 1 and \texttt{wtsum1(nretain)}. \ \texttt{over1(1}$)$ is
the first entry of \texttt{over1}. \ We note that \texttt{over1(1}$)$ must be less than or equal to \texttt{nretain}.
\ \texttt{seedvec} is the basis vector which corresponds with
\texttt{over1(1)}. \ \texttt{candidates} is a column vector consisting of all
basis vectors which have a non-zero matrix element with \texttt{seedvec}.
\ Since the task is to find new basis vectors, \texttt{seedvec} is removed
from \texttt{candidates}.

\texttt{weights2}\ is a column vector containing the absolute values of the
matrix transition elements from \texttt{seedvec}. \ \texttt{wtsum2} is an
ascending string between $0$ and 1 such that the difference between an entry
and its preceding entry is proportional to the value of \texttt{weights2}.
\ \ \texttt{over2} is a location list of elements of \texttt{wtsum2} greater
than a generated random number. \ \texttt{newvec} is the basis vector which
corresponds with \texttt{over2(1)}. \ \texttt{newvec} is appended to the set
\texttt{vecs}$.$ \ The function \texttt{unique} sorts the input list and
removes any repetitions. \ This loop is repeated until \texttt{vecs} has
\texttt{nactive} elements.\medskip

\noindent\textit{Redefine }$S$\textit{ as the subspace spanned by the updated
set of basis vectors and repeat steps 2 through 5.}

\noindent\qquad\texttt{for iter = 1:niter}

\noindent\qquad\qquad\texttt{...}

\noindent\qquad\texttt{end}

\noindent\qquad\texttt{save results.mat vecs v d}

\noindent The algorithm is iterated \texttt{niter} times and the final results
are saved in the file \texttt{results.mat}. \ More examples of QSE
diagonalization codes for various field theory systems written in MATLAB,
Fortran, and/or C++ will be made available in the future.

\section{Summary}

In this short tutorial we have listed some supplemental material for those
interested in writing or using codes that implement QSE methods. \ We have
discussed the calculation of Hamiltonian matrix elements and the selection of
new basis states. \ We have also presented a simple MATLAB program which
applies the QSE method to find the ground state of a matrix.

The discussion here was limited to the basic QSE method. \ As presented in the
Guangzhou workshop there have also been interesting new developments in QSE
diagonalization regarding the technique of stochastic error correction.
\ This technique will be presented in a forthcoming work \cite{sec}.

\section*{Acknowledgments}
I thank my collaborators on the papers cited here and the organizers and 
participants of the International Workshop on
Nonperturbative Methods and Lattice QCD in Guangzhou. \ Financial support was
provided by the National Science Foundation.

\end{document}